\documentclass[a4paper]{jpconf}
\usepackage{citesort,amsmath,graphicx,iopams}
\newcommand{\sig} 		{\sigma}
\newcommand{\kap} 		{\kappa}

\newcommand{\dSS} 	{\scriptscriptstyle}
\newcommand{\muB} 	{\mu_{\dSS B}}

\newcommand{\pdif} 	[1][]{\partial{#1}}                           

\begin{document}
\title{Susceptibilities from a black hole engineered EoS with a critical point}
\author{Israel Portillo}
\address{Department of Physics, University of Houston, Houston, TX 77204, USA}

\ead{iportillovazquez@gmail.com}

\begin{abstract}
Currently at the Beam Energy Scan at RHIC experimental efforts are being made to find the QCD critical point. On the theoretical side, the behavior of higher-order susceptibilities of the net-baryon charge from Lattice QCD at \(\muB\!=\!0\) may allow us to estimate the position of the critical point in the QCD phase diagram. However, even if the series expansion continues to higher-orders, there is always the possibility to miss the critical point behavior due to truncation errors.  An alternative approach is to use a black hole engineered holographic model, which displays a critical point at large densities and matches lattice susceptibilities at \(\muB\!=\!0\). Using the thermodynamic data from this black hole model, we obtain the freeze-out points extracted from the net-protons distribution measured at STAR and explore higher order fluctuations at the lowest energies at the beam energy scan to investigate signatures of the critical point.
\end{abstract}

\section{Introduction}

One of the most important goals of heavy ion physics is the mapping of the phases of strongly interacting matter at high temperatures, $T$, and densities. It is well established by lattice QCD calculations that at sufficiently high temperatures and zero baryonic chemical potential, $\muB$, there is a crossover between baryonic matter and a deconfined state of quarks and gluons~\cite{Aoki:2006we}. With increasing $\muB$, the crossover is expected to end in a critical point (CEP) that separates the crossover from a first order phase transition. This prominent feature of the QCD phase diagram is inferred from taking into account the effects of nonzero quark masses on chiral models \cite{Stephanov:1998dy}. 

The Beam Energy Scan (BES) at RHIC is exploring high density regions of the QCD phase diagram looking for experimental signatures of the CEP. One promising signature involves comparing moments of the distribution of the net protons compared to the susceptibilities of the pressure \cite{Karsch:2012wm}, which are expected to show non-monotonic behavior in the kurtosis \cite{Stephanov:2011pb}. 

While lattice QCD is unable to perform full  calculations at finite $\muB$ due to the Fermi sign problem, a small finite $\muB$ region can be reached employing different techniques such as Taylor expansions at $\muB\!=\!0$ \cite{Allton:2002zi,Gavai:2008zr} and analytic continuations from an imaginary chemical potential \cite{Wu:2006su,D'Elia:2002gd,deForcrand:2002hgr,Gunther:2016vcp}. Even if the CEP were located at an accessible density, it is possible that the non-monotonicity is missed by the extrapolation scheme, due to the singular behavior of the thermodynamic quantities close to the CEP. 

An alternative approach consists of studying the behavior of strongly interacting matter with a critical point using the holographic duality \cite{Maldacena:1997re}. Ref.\ \cite{Gubser:2008ny} showed how to construct black hole solutions of higher dimensional gravitational theories with thermodynamic properties similar to the QGP computed on the lattice at $\mu_B=0$. The generalization of this type of model to nonzero $\mu_B$ was done in \cite{DeWolfe:2010he,DeWolfe:2011ts}, where it was shown that these models can display a CEP at large baryon densities. These ``black hole engineered" non-conformal models possess a nonzero bulk viscosity \cite{Gubser:2008yx,Finazzo:2014cna}, which plays an important role in hydrodynamics simulations \cite{NoronhaHostler:2008ju,Noronha-Hostler:2013gga,Noronha-Hostler:2014dqa,Ryu:2015vwa}, and can be used to compute baryonic susceptibilities and transport coefficients at nonzero $\mu_B$ \cite{Rougemont:2015ona,Rougemont:2015wca}. 

Recently, the parameters in the effective dilaton potential of this black hole model were revised in \cite{Finazzo:2016mhm} to provide a better match to current lattice data at $\mu_B=0$. The details of the corresponding extension to $\mu_B\neq 0$ will be published elsewhere \cite{newpaper}. In these proceedings, we use the preliminary data from this revised black hole model to compare its baryonic susceptibilities with the fluctuations of net-protons \cite{Adamczyk:2013dal} and estimate the corresponding $(T,\muB)$ freeze-out line. 

\section{Baryon Number Fluctuations and the CEP}
The baryon number susceptibility is defined by the following derivatives of the pressure, $\chi_n(T,\muB)=\frac{\pdif^n}{\pdif(\muB/T)^n}\left(\frac{P}{T^4}\right)$, and these quantities can be numerically calculated in the black hole model at $\muB\neq 0$. Susceptibilities provide essential information about the effective degrees of freedom of a system, and are related directly to the moments of the distribution, from which volume-independent susceptibility ratios can be formed 
\begin{align}
\begin{aligned}
    	&&\text{mean : }&     &	           M &\;=\; \chi_1 &&&&&&
    	& M/\sig^2 		 &\;=\; \chi_1/\chi_2 &&\\
    	&&\text{variance : }&		& \sig^2 &\;=\; \chi_2 &&&&&&
    	& S\sig 			 	 &\;=\; \chi_3/\chi_2&&\\
    	&&\text{skewness : }&	&              S &\;=\; \chi_3/\chi_2^{3/2} &&&&&&
    	& \kap\sig^2  	 &\;=\; \chi_4/\chi_2&&\\
    	&&\text{kurtosis : }&		&     \kap  &\;=\; \chi_4/\chi_2^2 &&&&&&
    	& S\sig^3/M 		 &\;=\; \chi_3/\chi_1&&
\end{aligned}
\end{align}
In a heavy ion collision, the moments of the distribution are fixed at the chemical freeze-out such that comparisons to them may be used to extract $T$ and $\mu_B$ at freeze-out \cite{Borsanyi:2014ewa,Alba:2014eba,Noronha-Hostler:2016rpd}. Based on the singular behavior of thermodynamic variables close to the CEP in the theory of second order phase transitions, the susceptibilities scale with different powers of the (diverging) correlation length $\xi$ close to the CEP. For instance, $\chi_2 \sim VT\xi^2$ (at mean field level) where $V$ is the volume and, for a homogeneous system in equilibrium, one can show that the high order susceptibilities diverge with higher powers of $\xi$ \cite{Stephanov:2008qz}. In practice, the divergence of $\xi$ is limited by the system size and by finite time effects. In Ref.\ \cite{Stephanov:2011pb} it was argued that the characteristic behavior of the ratio \(\kap\sig^2\) close to the critical point has a non-monotonic dependence as one approaches the CEP.

\section{Results}

\begin{figure}[h]
\begin{tabular}{cc}
\includegraphics[width=19pc]{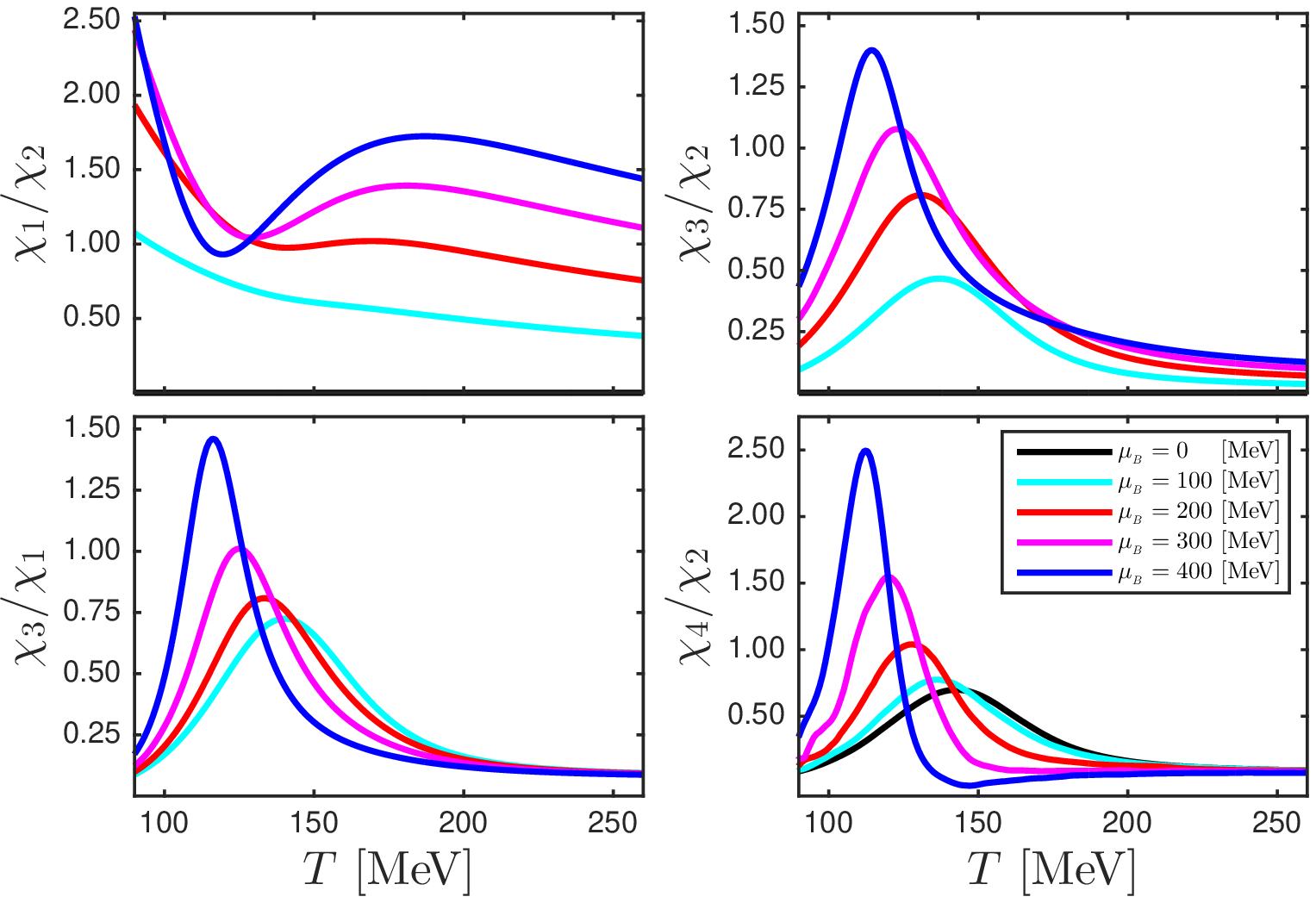} & \includegraphics[width=19pc]{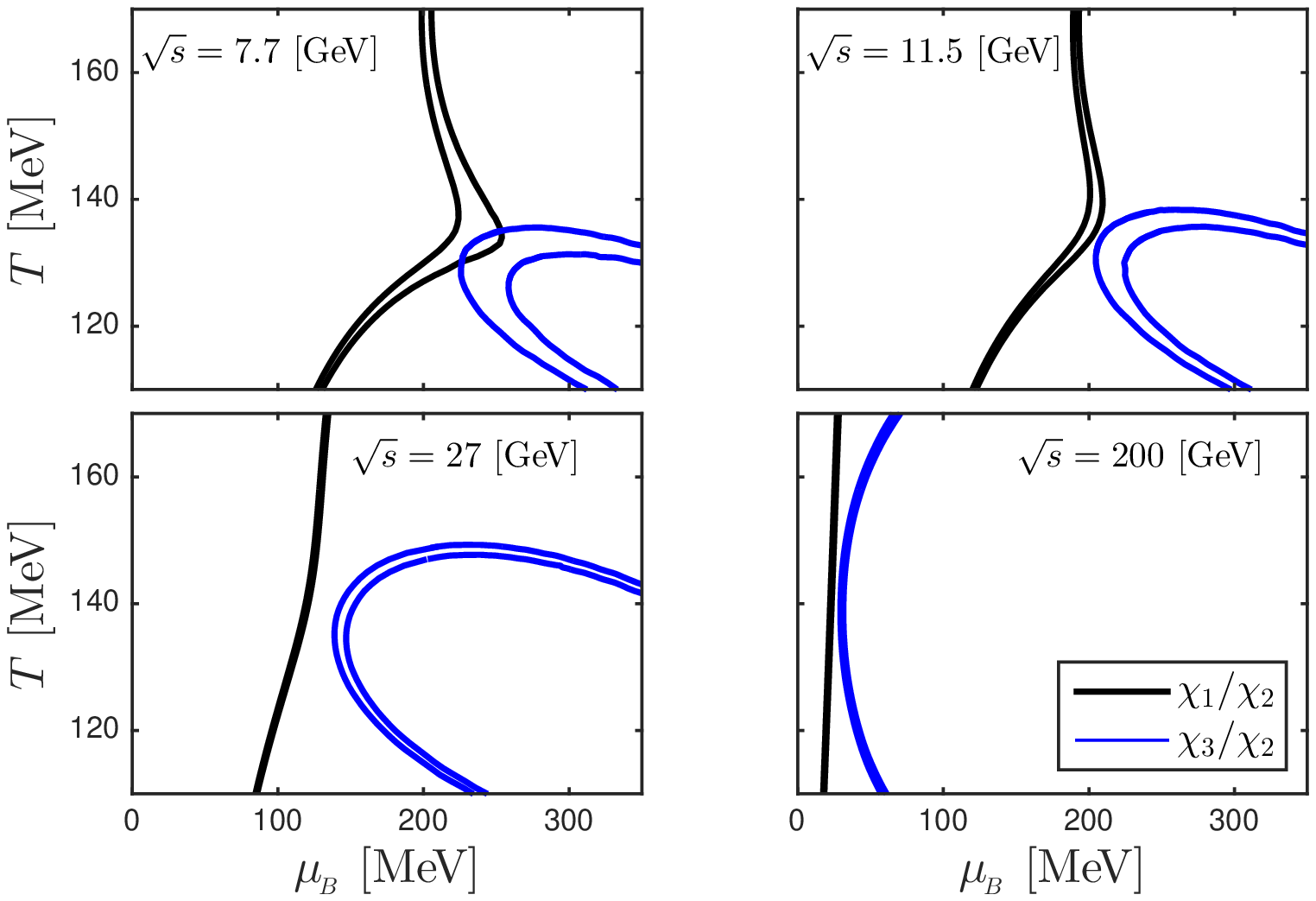}
\end{tabular}
	\caption{\label{Fig:Rat}(Color online) Susceptibility ratios obtained from the black hole model (left panel) for $\mu_B=0$ (black lines), $\mu_B=100$ MeV (cyan lines), $\mu_B=200$ MeV (red lines), $\mu_B=300$ MeV (magenta lines), and $\mu_B=400$ MeV (blue lines). Trajectories in the (\(T,\mu_B\)) plane (right panel) that satisfy the STAR data~\cite{Adamczyk:2013dal} as a function of $\sqrt{s}$.}
\end{figure}

In Fig.~\ref{Fig:Rat} (left) the susceptibility ratios calculated in the model are shown as a function of $T$ for different values of $\muB$. From these curves, we were able to extract the bands in $T$ and $\muB$ for each collision energy $\sqrt{s}=7.7-200$ GeV in Fig.~\ref{Fig:Rat} (right), by imposing that they reproduce the corresponding $\chi_1/\chi_2$ and $\chi_3/\chi_2$ from STAR. While some energies show a clear overlap, some of the lower energies do not, which may be due to other effects such as decays or acceptance cuts. In those cases, the freeze-out is obtained from the middle point of the overlapping area. For the rest of the energies, we choose from each area the point that gives the smallest distance of one area with respect to the other one. The freeze-out points, with their respective error bars, are extracted from the gap between those two points.

\begin{figure}[h]
	\begin{center}
		\includegraphics[width=25pc]{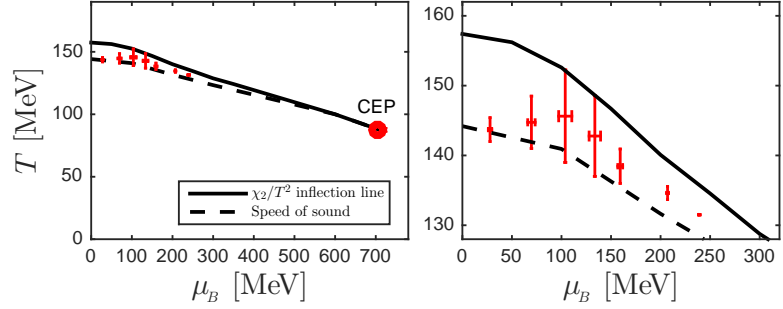}
		\caption{\label{Fig:RatPhDia}(Color online) Freeze-out line (red data points) extracted by comparing the model calculations to STAR data \cite{Adamczyk:2013dal}. The preliminary estimate for the location of the black hole model's CEP is shown. The locations of the minimum of $c_s^2$ (dashed line) and the inflection point of $\chi_2/T^2$ (solid line) of the black hole model \cite{newpaper} are also shown for a comparison.}
	\end{center}
\end{figure}

Fig.\ \ref{Fig:RatPhDia} shows an estimate for the location of the CEP found in the model, as well as our results for the $(T,\muB)$ freeze-out points. It is interesting to note that they are placed between the line corresponding to the inflection point of $\chi_2/T^2$ (solid line) and the one corresponding to the minimum of $c_s^2$ (dashed line), computed using the revised black hole model at nonzero $\mu_B$ \cite{newpaper}.

\begin{figure}[h]
	\begin{center}
		\includegraphics[width=35pc]{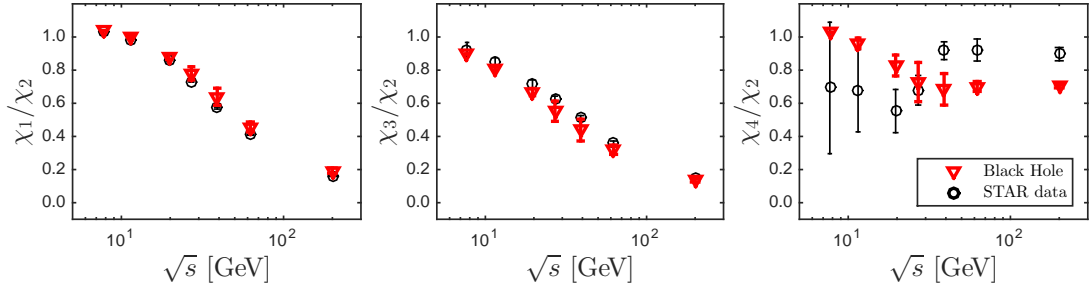}
	\end{center}
    \caption{\label{Fig:Pred}(Color online) Low order baryon susceptibility ratios, $\chi_1/\chi_2$ and $\chi_3/\chi_2$, used to obtain the freeze-out line compared to the net-proton distribution from STAR \cite{Adamczyk:2013dal}. Also shown is the $\chi_4/\chi_2$ ratio from STAR \cite{Adamczyk:2013dal} (black points) together with the black hole model calculations (red arrows) along the freeze-out line.}
\end{figure}

In Fig.\ \ref{Fig:Pred}, model calculations for $\chi_1/\chi_2$ and $\chi_3/\chi_2$ are compared to their corresponding experimental data points from STAR \cite{Adamczyk:2013dal}. We also show in Fig.\ \ref{Fig:Pred} the $\chi_4/\chi_2$ ratio computed at the extracted freeze-out $(T,\mu_B)$ values. One can see that this ratio increases with decreasing $\sqrt{s}$, following the trend of current experimental data at low $\sqrt{s}$, even though the CEP of the model only appears at much larger densities. 
		
\section{Conclusions}
We used preliminary data from the black hole engineered holographic model \cite{newpaper} to extract the freeze-out line across collision energies at the BES. The computed $\chi_4/\chi_2$ ratio along the freeze-out line grows with decreasing $\sqrt{s}$, even though the freeze-out values are still far from the high density region where the CEP of the model is located \cite{newpaper}. Preliminary STAR data with higher $p_T$ cuts ($0.4<p_T<2.0$ GeV) \cite{Luo:2015ewa} show larger enhancement for $\chi_4/\chi_2$ at low $\sqrt{s}$ and we plan to examine this new data taking into account effects from acceptance cuts and decays in the future.  
\section*{References}
\bibliography{Isra_HQ16}{}

\providecommand{\newblock}{}
\begin{thebibliography}{10}
\expandafter\ifx\csname url\endcsname\relax
  \def\url#1{{\tt #1}}\fi
\expandafter\ifx\csname urlprefix\endcsname\relax\def\urlprefix{URL }\fi
\providecommand{\eprint}[2][]{\url{#2}}

\bibitem{Aoki:2006we}
Aoki Y, Endrodi G, Fodor Z, Katz S~D and Szabo K~K 2006 {\em Nature\/} {\bf
  443} 675--678

\bibitem{Stephanov:1998dy}
Stephanov M~A, Rajagopal K and Shuryak E~V 1998 {\em Phys. Rev. Lett.\/} {\bf
  81} 4816--4819

\bibitem{Karsch:2012wm}
Karsch F 2012 {\em Central Eur. J. Phys.\/} {\bf 10} 1234--1237

\bibitem{Stephanov:2011pb}
Stephanov M~A 2011 {\em Phys. Rev. Lett.\/} {\bf 107} 052301

\bibitem{Allton:2002zi}
Allton C~e~a 2002 {\em Phys. Rev.\/} {\bf D66} 074507 (\textit{Preprint}
  \eprint{hep-lat/0204010})

\bibitem{Gavai:2008zr}
Gavai R~V and Gupta S 2008 {\em Phys. Rev.\/} {\bf D78} 114503
  (\textit{Preprint} \eprint{0806.2233})

\bibitem{Wu:2006su}
Wu L~K, Luo X~Q and Chen H~S 2007 {\em Phys. Rev.\/} {\bf D76} 034505
  (\textit{Preprint} \eprint{hep-lat/0611035})

\bibitem{D'Elia:2002gd}
D'Elia M and Lombardo M~P 2003 {\em Phys. Rev.\/} {\bf D67} 014505
  (\textit{Preprint} \eprint{hep-lat/0209146})

\bibitem{deForcrand:2002hgr}
de~Forcrand P and Philipsen O 2002 {\em Nucl. Phys.\/} {\bf B642} 290--306
  (\textit{Preprint} \eprint{hep-lat/0205016})

\bibitem{Gunther:2016vcp}
Gunther J, Bellwied R, Borsanyi S, Fodor Z, Katz S~D, Pasztor A and Ratti C
  2016  (\textit{Preprint} \eprint{1607.02493})

\bibitem{Maldacena:1997re}
Maldacena J~M 1999 {\em Int. J. Theor. Phys.\/} {\bf 38} 1113--1133 [Adv.
  Theor. Math. Phys.2,231(1998)]

\bibitem{Gubser:2008ny}
Gubser S~S and Nellore A 2008 {\em Phys. Rev.\/} {\bf D78} 086007
  (\textit{Preprint} \eprint{0804.0434})

\bibitem{DeWolfe:2010he}
DeWolfe O, Gubser S~S and Rosen C 2011 {\em Phys. Rev.\/} {\bf D83} 086005
  (\textit{Preprint} \eprint{1012.1864})

\bibitem{DeWolfe:2011ts}
DeWolfe O, Gubser S~S and Rosen C 2011 {\em Phys. Rev.\/} {\bf D84} 126014
  (\textit{Preprint} \eprint{1108.2029})

\bibitem{Gubser:2008yx}
Gubser S~S, Nellore A, Pufu S~S and Rocha F~D 2008 {\em Phys. Rev. Lett.\/}
  {\bf 101} 131601 (\textit{Preprint} \eprint{0804.1950})

\bibitem{Finazzo:2014cna}
Finazzo S~I, Rougemont R, Marrochio H and Noronha J 2015 {\em JHEP\/} {\bf 02}
  051 (\textit{Preprint} \eprint{1412.2968})

\bibitem{NoronhaHostler:2008ju}
Noronha-Hostler J, Noronha J and Greiner C 2009 {\em Phys. Rev. Lett.\/} {\bf
  103} 172302 (\textit{Preprint} \eprint{0811.1571})

\bibitem{Noronha-Hostler:2013gga}
Noronha-Hostler J, Denicol G~S, Noronha J, Andrade R~P~G and Grassi F 2013 {\em
  Phys. Rev.\/} {\bf C88} 044916

\bibitem{Noronha-Hostler:2014dqa}
Noronha-Hostler J, Noronha J and Grassi F 2014 {\em Phys. Rev.\/} {\bf C90}
  034907 (\textit{Preprint} \eprint{1406.3333})

\bibitem{Ryu:2015vwa}
Ryu S, Paquet J~F, Shen C, Denicol G~S, Schenke B, Jeon S and Gale C 2015 {\em
  Phys. Rev. Lett.\/} {\bf 115} 132301

\bibitem{Rougemont:2015ona}
Rougemont R, Noronha J and Noronha-Hostler J 2015 {\em Phys. Rev. Lett.\/} {\bf
  115} 202301 (\textit{Preprint} \eprint{1507.06972})

\bibitem{Rougemont:2015wca}
Rougemont R, Ficnar A, Finazzo S and Noronha J 2016 {\em JHEP\/} {\bf 04} 102
  (\textit{Preprint} \eprint{1507.06556})

\bibitem{Finazzo:2016mhm}
Finazzo S~I, Critelli R, Rougemont R and Noronha J 2016 {\em Phys. Rev.\/} {\bf
  D94} 054020 (\textit{Preprint} \eprint{1605.06061})

\bibitem{newpaper}
Finazzo S~I, Critelli R, Rougemont R and Noronha J 2016, to appear

\bibitem{Adamczyk:2013dal}
Adamczyk L {\em et~al.\/} (STAR) 2014 {\em Phys. Rev. Lett.\/} {\bf 112} 032302

\bibitem{Borsanyi:2014ewa}
Borsanyi S, Fodor Z, Katz S~D, Krieg S, Ratti C and Szabo K~K 2014 {\em Phys.
  Rev. Lett.\/} {\bf 113} 052301

\bibitem{Alba:2014eba}
Alba P, Alberico W, Bellwied R, Bluhm M, Mantovani~Sarti V, Nahrgang M and
  Ratti C 2014 {\em Phys. Lett.\/} {\bf B738} 305--310 (\textit{Preprint}
  \eprint{1403.4903})

\bibitem{Noronha-Hostler:2016rpd}
Noronha-Hostler J, Bellwied R, Gunther J, Parotto P, Pasztor A, Vazquez I~P and
  Ratti C 2016  (\textit{Preprint} \eprint{1607.02527})

\bibitem{Stephanov:2008qz}
Stephanov M~A 2009 {\em Phys. Rev. Lett.\/} {\bf 102} 032301 (\textit{Preprint}
  \eprint{0809.3450})

\bibitem{Luo:2015ewa}
Luo X (STAR) 2015 {\em PoS\/} {\bf CPOD2014} 019 (\textit{Preprint}
  \eprint{1503.02558})

\end{thebibliography}
\bibliographystyle{nucl}

\end{document}